\documentclass[aps,prb,twocolumn,groupedaddress,floats,showpacs]{revtex4}

\usepackage{amsmath}
\usepackage{amssymb}
\usepackage{graphicx}

\newcommand{\e}{\varepsilon}

\newcommand{\pot}{{\mu}}

\newcommand{\up}{\uparrow}
\newcommand{\dw}{\downarrow}

\renewcommand{\vec}[1]{\mathbf{#1}}
\renewcommand{\vr}{\vec{r}}
\newcommand{\vm}{\vec{m}}
\newcommand{\ve}{\vec{e}}
\newcommand{\vsigma}{\mbox{\boldmath $\sigma$}}

\newcommand{\vnabla}{\mbox{\boldmath $\nabla$}}
\newcommand{\vk}{\vec{k}} 
 
\newcommand{\vkp}{\vec{k}_{\parallel}} 
\newcommand{\vrp}{\vec{r}_{\parallel}}

\begin{document}

\title{Impurity-assisted Andreev reflection at a spin-active half-metal--superconductor interface}

\author{Francis B. Wilken}

\author{Piet W. Brouwer}

\affiliation{Dahlem Center for Complex Quantum Systems and Institut f\"ur Theoretische Physik, Freie Universit\"at Berlin, Arnimallee 14, 14195 Berlin, Germany}
\date{\today}
\begin{abstract}
Andreev reflection at a half-metal--superconductor interface.
\end{abstract}


\begin{abstract} 
The Andreev reflection amplitude at a clean interface between a half-metallic ferromagnet (H) and a superconductor (S) for which the half metal's magnetization has a gradient perpendicular to the interface is proportional to the excitation energy $\varepsilon$ and vanishes at $\varepsilon=0$ [B\'{e}ri {\em et al.}, Phys.\ Rev.\ B {\bf 79}, 024517 (2009)]. Here we show that the presence of impurities at or in the immediate vicinity of the HS interface leads to a finite Andreev reflection amplitude at $\varepsilon=0$. This impurity-assisted Andreev reflection dominates the low-bias conductance of a HS junction and the Josephson current of an SHS junction in the long-junction limit.
\end{abstract}
\maketitle

\section{Introduction}

The experimental observation of a sizable supercurrent through a Josephson junction containing the half-metallic ferromagnet CrO$_2$ \cite{kn:keizer2006,kn:anwar2010,kn:anwar2011} has renewed theoretical interest in the superconductor proximity effect in half metals.\cite{kn:asano2007a,kn:asano2007b,kn:eschrig2008,kn:beri2009,kn:kupferschmidt2009,kn:linder2010,kn:kupferschmidt2011,kn:duckheim2011,kn:chung2011} Because of the absence of minority spin carriers in a half metal,\cite{kn:lewis1997,kn:coey2002} the induced superconducting correlations must be of the spin-triplet type, even if they arise from proximity to a spin-singlet superconductor.\cite{kn:bergeret2001,kn:bergeret2001b,kn:kadigrobov2001,kn:eschrig2003,kn:bergeret2005} 

The spin-triplet proximity effect is mediated by Andreev reflections\cite{kn:andreev1964} of majority electrons into majority holes. Since such Andreev reflections violate spin conservation --- spin-conserving Andreev reflection reflects majority electrons into minority holes ---, they require breaking of the spin-rotation symmetry around the half metal's spin quantization axis. A prominent mechanism for the breaking of the spin-rotation symmetry is a magnetization gradient perpendicular to the interface between the superconductor and the half metal, see Fig.\ \ref{fig:1}, which occurs naturally in the presence of a magnetic interface anisotropy that favors a different magnetization direction than the magnet's bulk anisotropy.\cite{kn:ohandley2000} In the literature, an interface with a different magnetization direction than the bulk magnetization direction is referred to as ``spin active''.\cite{kn:eschrig2008} Artificially created spin active interfaces have been shown to be responsible for the observation of the spin-triplet proximity effect in standard (non half-metallic) ferromagnets.\cite{kn:khaire2010,kn:klose2011,kn:wang2011} In view of the observed bi-axial magnetic anisotropy of CrO$_2$ thin films,\cite{kn:miao2005,kn:keizer2006,kn:goennenwein2007} spin-active interface have also been proposed as an explanation for the observed Josephson effect in the CrO$_2$-based Josephson junctions,\cite{kn:eschrig2008} but there is no direct experimental evidence confirming this link.\cite{kn:keizer2006}

What makes the spin-triplet proximity effect in a half metal particularly interesting from a theoretical point of view, is that there is a symmetry argument that strongly limits the effectiveness of a spin-active interface as a source of spin-triplet superconducting correlations.\cite{kn:beri2009,kn:kupferschmidt2011} This symmetry argument applies to the case of a half metal, but not to a standard ferromagnet (with incomplete spin polarization). It essentially forbids Andreev reflection of majority electrons into majority holes or vice versa if five key conditions are met:\cite{footn1} Particle-hole symmetry, translation symmetry along the superconductor interface, $\pi$-rotation symmetry around an axis normal to the superconductor interface, quasiparticle conservation, and the absence of minority carriers. All five conditions are met for Andreev reflection of carriers at the Fermi level in a clean spin-active interface between a half metal (H) and a superconductor (S). Away from the Fermi level, particle-hole degeneracy is lifted, and generically the Andreev reflection amplitudes are proportional to the excitation energy $\varepsilon$.\cite{kn:beri2009,kn:kupferschmidt2011,kn:eschrig2009} This energy dependence leads to an interface conductance $G \propto (e V)^2$, where $V$ is the applied bias, and to a Josephson current $I \propto \max(k_{\rm B} T,E_L)^3$ for a clean SHS junction at temperature $T$ in the ``long junction'' limit (gap $\Delta_0$ much larger than the junction's Thouless energy $E_L$).\cite{kn:beri2009,kn:kupferschmidt2011,kn:eschrig2009} In contrast, for junctions involving a standard ferromagnet, the zero-bias conductance $G$ is finite, whereas $I \propto \max(k_{\rm B} T,E_L)$.\cite{kn:dejong1995,kn:bergeret2005,kn:houzet2007,kn:volkov2010}

In order to open up the possibility of Andreev reflection at the Fermi level, one of the remaining four conditions listed above has to be lifted. Several options have been investigated in the literature, and have been shown to lead to a finite zero-bias conductance $G$ of an HS junction and to a temperature dependence of the Josephson current of a ballistic SHS junction that closely resembles junctions with a standard ferromagnet instead of a half-metal. These include the breaking of the rotation symmetry around an axis normal to the superconductor interface by a magnetization gradient parallel to the interface\cite{kn:kupferschmidt2009,kn:kupferschmidt2011} or by spin-orbit coupling in the superconductor,\cite{kn:duckheim2011} or the inelastic scattering of quasiparticles in the half metal, which effectively lifts quasiparticle conservation.\cite{kn:beri2009b}

In this article, we investigate the lifting of the translation symmetry along the interface by scattering from non-magnetic impurities. This is particularly relevant for CrO$_2$-based junctions, as interfaces between the metastable compound CrO$_2$ and other materials are notoriously poorly defined.\cite{kn:coey2002} We confirm that impurity scattering, too, is a viable mechanism for Andreev reflection of majority electrons into majority holes at a HS junction at the Fermi level. Moreover, we show that this impurity-assisted Andreev reflection remains coherent, with a well-defined magnitude and phase, after performing an average over different impurity configurations. As a result, impurity-assisted Andreev reflection not only gives rise to a finite zero-bias conductance, but also to a strong enhancement of the Josephson current at low temperatures. Our analytical calculations are consistent with a numerical analysis of the triplet proximity effect in a strongly disordered half metal by Asano {\em et al.},\cite{kn:asano2007a,kn:asano2007b} who did not report any suppression of the induced superconducting correlations for energies near the Fermi level.

The detailed outline of this article is as follows: In section \ref{sec:HS-interface}, we describe the model Hamiltonian used in our calculations and review the basic symmetry relations for the scattering matrix of an HS interface. Then, in Sec.\ \ref{sec:perturbation} we calculate the Andreev amplitude for an HS interface with a single impurity, up to quadratic order in the impurity potential. We apply our results to an interface with a finite density of impurities in Sec.\ \ref{sec:applications}, where we show that the presence of impurities at the interface leads to a finite interface conductance at zero bias and to a significantly enhanced Josephson current at low temperatures. We conclude in Sec.\ \ref{sec:summary}. The appendices contain various additional results for scenarios not covered in the main text.

\section{Half-metal--superconductor interface with perpendicular magnetization gradient}\label{sec:HS-interface}

\subsection{Model Hamiltonian without perturbations}\label{sec:hamilton-operator} \label{sec:2a}

\begin{figure}
  \centering
  \includegraphics[width = 0.95 \linewidth]{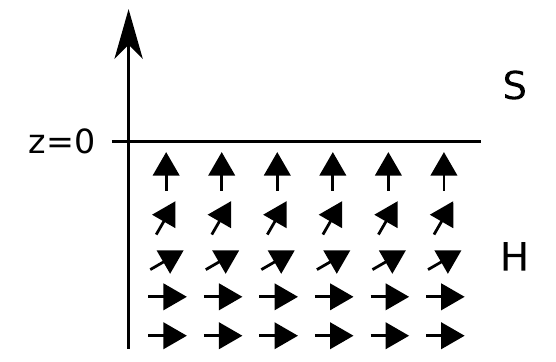}
  \caption{HS interface with a spin-active interface, originating from a magnetization gradient perpendicular to the interface. Arrows indicate the magnetization direction.}
  \label{fig:1}
\end{figure}

Our calculation builds on the calculation of the Andreev reflection amplitudes for a clean half-metal--superconductor interface with a perpendicular magnetization gradient by Kupferschmidt and one of the authors.\cite{kn:kupferschmidt2011} Following Ref.\ \onlinecite{kn:kupferschmidt2011}, we choose coordinates, such that the superconductor occupies the half space $z > 0$ and the HS interface is in the $xy$ plane, see Fig.\ \ref{fig:1}. Periodic boundary conditions are applied in the $x$ and $y$ directions, with periods $W_x$ and $W_y$, respectively. Quasiparticle excitations near the HS interface are described by the Bogoliubov-de Gennes equation\cite{kn:beenakker1995}
\begin{equation} 
\label{eq:BdG}
  {\cal H}
 \Psi(\vr)
  = \e \Psi(\vr),
  \ \ 
  {\cal H} = {\cal H}_0 + {\cal V},
\end{equation}
where
\begin{equation}
  {\cal H}_0 = \left( \begin{array}{cc} 
  \hat H_0   &   i \Delta(\vr) \sigma_2 \\
- i \Delta^*(\vr) \sigma_2 & - \hat H_0^*  
\end{array} \right),
\end{equation}
and the four-component spinor
\begin{equation} 
 \Psi(\vr)  = ( u_{\up }(\vr) , u_{\dw}(\vr) , v_{\up}(\vr) , v_{\dw}(\vr) )^{\rm T}, \label{eq:psispinor}
\end{equation}
consists of wavefunctions $u_{\sigma}(\vr) $ for the electron and $v_{\sigma}(\vr)$ for the hole degrees of freedom. The $4 \times 4$ matrix operator ${\cal H}_0$ describes the HS junction in the absence of a magnetization gradient and impurity scattering. These two effects are described by the perturbation ${\cal V}$ and will be discussed in the next subsection.

The superconducting order parameter ${\Delta (\vr)=\Delta_0 e^{i \phi} \Theta(z)}$, where $\Theta(z) = 1$ if $z > 0$ and $0$ otherwise. This step function model is a good approximation for tunnelling interfaces of $s$-wave superconductors.\cite{kn:likharev1979} For the single-particle Hamiltonian $\hat H_0$ we take
\begin{equation}
  \hat H_0 =
  -\frac{\hbar^2}{2 m} \vnabla^2 -
  \sum_{\sigma} \pot_{\sigma}(z) \hat P_{\sigma} + \hbar w \delta(z),
  \label{eq:Hnormal}
\end{equation}
where $m$ is the effective electron mass (taken to be equal on both sides of the interface), and 
\begin{equation}
  \pot_{\sigma}(z) = \left\{ \begin{array}{ll}
  \pot_{{\rm H}\sigma} & \mbox{if $z < 0$}, \\
  \pot_{{\rm S}} & \mbox{if $z > 0$}, \end{array} \right.
\end{equation}
with $\sigma=\uparrow,\downarrow$ and the potentials $\pot_{{\rm H}\uparrow}$, $\pot_{{\rm H}\downarrow}$, and $\pot_{\rm S}$ representing the combined effect of the chemical potential and band offsets for the majority and minority electrons in the half metal and for the superconductor, respectively, and where $w$ sets the strength of a delta-function potential barrier at the interface. The operators 
\begin{equation}
  \hat P_{\uparrow} = \frac{1}{2} + \frac{1}{2} \ve_3 \cdot \hat{\vsigma}, \;
  \hat P_{\downarrow} = \frac{1}{2} - \frac{1}{2} \ve_3 \cdot \hat{\vsigma},
\end{equation}
project onto the majority and minority components, respectively, where we have taken $\ve_3$ to be the unit vector pointing along the magnetization direction in the half metal.

The potentials $\pot_{{\rm H}\uparrow}$, $\pot_{{\rm H}\downarrow}$, and $\pot_{\rm S}$ are such that $\pot_{{\rm H}\uparrow}$, $\pot_{{\rm S}} > 0$, and $\pot_{{\rm H}\downarrow} < 0$. As a result, majority states in the half metal with uniform $\vm(\vr)$ and states in the normal state of the superconductor are propagating states, with Fermi wavenumbers 
\begin{equation}
  k_{\uparrow} = \frac{1}{\hbar}\sqrt{2 m \pot_{{\rm H}\uparrow}},\ \
  k_{\rm S} = \frac{1}{\hbar}\sqrt{2 m \pot_{\rm S}}, \label{eq:k-approx}
\end{equation}
respectively. The corresponding Fermi velocities are ${v_{\uparrow} = \hbar k_{\uparrow}/m}$ and ${v_{\rm S} = \hbar k_{\rm S}/m}$, respectively.
Minority states in the half metal are evanescent with wavefunction decay rate
\begin{equation}
  \kappa_{\downarrow} = \frac{1}{\hbar} \sqrt{2 m |\pot_{{\rm H}\downarrow}|}.
  \label{eq:kappadown}
\end{equation}
The Andreev approximation $\Delta_0 \ll \mu_{\rm S}$ is used throughout our calculation.

The propagation of electrons with Hamiltonian (\ref{eq:BdG}) is described by the $4 \times 4$ matrix Green function ${\cal G}(\varepsilon;\vr,\vr')$, which is a solution of the Gorkov equation
\begin{equation}
  (\varepsilon - {\cal H}) {\cal G}(\varepsilon;\vr,\vr') = \delta(\vr-\vr').
  \label{eq:Gorkov}
\end{equation}
The Green function can be written
\begin{equation}
  {\cal G}(\varepsilon;\vr,\vr') = \frac{1}{W_x W_y}
    \sum_{\vk_{\parallel}} {\cal G}_z(\varepsilon;\vk_{\parallel};z,z')
  e^{i \vkp \cdot (\vrp-\vrp')}, \label{eq:G01}
\end{equation}
where $\vk_{\parallel} = k_x \ve_x + k_y \ve_y$ is a wavevector parallel to the superconductor interface and $\vrp = x \ve_x + y \ve_y$.\cite{footn2}
Solution of Eq.\ (\ref{eq:Gorkov}) then gives the result
\begin{equation}
  {\cal G}_z(\varepsilon;\vkp;z,z') = \frac{1}{i \hbar} (U_{\uparrow},U_{\downarrow},V_{\uparrow},V_{\downarrow}),
\end{equation}
where the first two column vectors $U_{\uparrow}$ and $U_{\downarrow}$ have the form
\begin{eqnarray}
  U_{\uparrow} &=& \frac{1}{v_{\uparrow z}}
  \left( \begin{array}{c}
  e^{i k_{\uparrow z}(\varepsilon)|z-z'|} + \rho_{\up}(\varepsilon) e^{-i k_{\uparrow z}(\varepsilon)(z+z')} \\
  0 \\
  0 \\
  \rho_{\dw}(-\varepsilon) e^{-i k_{\uparrow z}(\varepsilon) z' + \kappa_{\dw z}(-\varepsilon) z} \end{array} \right), \nonumber \\
  U_{\downarrow} &=&
  \frac{1}{i v_{\downarrow z}}
  \left( \begin{array}{c} 0 \\
  e^{-\kappa_{\dw z}(\varepsilon)|z-z'|} + \tau_{\dw}(\varepsilon) e^{\kappa_{\dw z}(\varepsilon)(z+z')} \\
  \tau_{\up}(-\varepsilon) e^{i k_{\uparrow z}(-\varepsilon)  z + \kappa_{\dw z}(\varepsilon) z'} \\ 0 \end{array} \right), \nonumber \\ \label{eq:UU}
\end{eqnarray}
for $z$, $z' < 0$, whereas the latter two column vectors $V_{\uparrow}$ and $V_{\downarrow}$ are obtained from $U_{\uparrow}$ and $U_{\downarrow}$, respectively, by particle-hole conjugation (complex conjugation, interchange of first and third, and of second and fourth rows, and the replacements $\varepsilon \to -\varepsilon$ and $\vkp \to - \vkp$, see App.\ \ref{sec:appa}). Here 
\begin{eqnarray}
  k_{\uparrow z}(\varepsilon) &=& \sqrt{k_{\uparrow}^2 - k_{\parallel}^2} + \varepsilon/\hbar v_{\uparrow z}, \label{eq:kup} \nonumber \\
  \kappa_{\downarrow z}(\varepsilon) &=& \sqrt{\kappa_{\dw}^2 + k_{\parallel}^2} - \varepsilon/\hbar v_{\downarrow z}, \label{eq:kdown}
\end{eqnarray}
up to corrections of order $\varepsilon^2$, with ${v_{\uparrow z} = \hbar k_{\uparrow z}(0)/m}$ and ${v_{\downarrow z} = \hbar k_{\downarrow z}(0)/m}$. The coefficient $\rho_{\up}(\varepsilon)$ is the amplitude for normal reflection at the HS interface in the absence of the perturbations, see Eq.\ (\ref{eq:S0}) below. Since there is no Andreev reflection in the absence of a magnetization gradient, one has $|\rho(\varepsilon)| = 1$. The other coefficients do not have an interpretation in terms of scattering amplitudes. Detailed expressions for the coefficients $\rho_{\sigma}$ and $\tau_{\sigma}$ can be found in App.\ \ref{sec:appa}.

In the normal state (for $\Delta_0 = 0$), majority electrons incident onto the superconductor interface are transmitted with probability
\begin{eqnarray}
  T_{\uparrow}(\vkp) &=&
  \frac{4 v_{\uparrow z} v_{\rm S z}}
    {4 w^2 + (v_{\uparrow z} + v_{{\rm S}z})^2},
\end{eqnarray}
where $v_{{\rm S}z} = (\hbar/m)(k_{\rm S}^2 - k_{\parallel}^2)^{1/2}$.
The use of the step-function model for the superconducting order parameter used in Eq.\ (\ref{eq:BdG}) requires that the normal-state transparency of the HS interface is small for majority electrons as well as minority electrons.\cite{kn:likharev1979} In our model, this condition is met if 
\begin{equation}
  \max(|w|,v_{{\rm S}}) \gg \max(v_{\uparrow},v_{\downarrow}),
\end{equation}
and we will assume that this inequality is met in the results we present below. In our final results, we will eliminate $w$ and $v_{\rm S}$ in favor of the normal-state reflection amplitude $r$ and the transmission coefficient $T_{\uparrow}(\vk)$ given above. Expressions for the case of an ideal interface with transparency $T_{\uparrow}(\vkp) = 1$, but still using the step-function model for $\Delta(\vr)$ at the interface, are given in App.\ \ref{sec:appendixideal}.

\subsection{Perturbation: Changing magnetization and impurities}

The perturbation ${\cal V}$ describes the combined effect of a magnetization direction that slowly changes as a function of the coordinate $z$, as well as impurity scattering. We take the magnetization direction to be\cite{kn:kupferschmidt2011}
\begin{equation}
\vec{m} = 
 (\ve_1 \cos \varphi +  \ve_2 \sin\varphi)\sin \theta(\vec{r})
  + \ve_3 \cos \theta(\vec{r}) ,
\end{equation}
where the polar angle
\begin{equation}
\theta(z) = \left\{ \begin{matrix} z / l_\text{d} & \text{if } z < 0, \\ 0 & \text{if } z \geq 0, \end{matrix} \right.
\end{equation}
while the azimuthal angle $\varphi$ remains constant. Transforming to a spin coordinate system in which $\vm$ points along $\ve_3$ everywhere in space and expanding up to first order in the gradient $d\theta/dz$, one finds that the single-particle Hamiltonian $\hat H$ takes the form $\hat H_0$ described in the previous subsection, with the additional term\cite{kn:volovik1987}
\begin{eqnarray}
  \hat V_{\rm m} &=&
  \frac{i \hbar}{2 m} (\sigma_2 \cos \varphi - \sigma_1 \sin \varphi)
  \left(
  \frac{d\theta}{dz} \frac{\partial}{\partial z} +
  \frac{\partial}{\partial z} \frac{d\theta}{dz} \right) \nonumber \\ &=&
  \frac{i \hbar}{m l_{\rm d}} (\sigma_2 \cos \varphi - \sigma_1 \sin \varphi)
  \left(\frac{\partial}{\partial z} - \frac{1}{2} \delta(z) \right).
\end{eqnarray}
The impurity potential is taken to be a sum of the form
\begin{equation}
  \hat V_{\rm i} = \sum_{j} u_{j} \delta(\vr-\vr_j),
  \label{eq:vimpurity}
\end{equation}
where $\vr_j$ is the position of the $j$th impurity and $u_{j}$ its strength. The impurity strength $u_{j}$ is related to the scattering cross section $\sigma_j$ for majority electrons,
\begin{equation}
  \sigma_j = \frac{m^2 u_{j}^2}{\pi \hbar^4}. \label{eq:sigma}
\end{equation}
In the main text, we will consider the case that the impurities are randomly positioned with density $n_{\rm imp}$ in the vicinity of the interface. Here it is important to point out, that $n_{\rm imp}$ is the impurity density at the interface, and that there may be a different impurity density in the bulk of the half metal. (Our results below show that only impurities within a distance $\sim 1/\kappa_{\dw}$ from the HS interface contribute to Andreev reflection. Hence, only the impurity density near the interface enters in our final expressions.) The situation that all impurities are located precisely at the interface is discussed separately in appendix \ref{sec:appendix0}.

Combining these two perturbations, we thus find
\begin{equation}
  {\cal V} = \left(\begin{matrix} \hat V_{\rm m} + \hat V_{\rm i} & 0 \\ 0 & -\hat V_{\rm m}^* -\hat V_{\rm i}^\star \end{matrix}\right).
\end{equation}

\subsection{Scattering matrix}

There are two linearly independent solutions of the Bogoliubov-de Gennes equation (\ref{eq:BdG}) for each wavevector $\vkp$. At a large distance from the HS interface ($z \ll 0$ for the coordinate system used here), they can be taken to be of the standard form
\begin{eqnarray}
  \label{eq:Psistandarde}
  \Psi_{\varepsilon\vkp{\rm e}}(\vr) &=&
  \frac{e^{i \vkp \cdot \vrp + i k_{\up z}(\varepsilon) z}}{\sqrt{v_{\uparrow z}(\varepsilon) W_x W_y}}
  \left(  \begin{array}{c} 1 \\ 0 \\ 0 \\ 0 \end{array} \right)
  \\ && \nonumber \mbox{}
  +
  \sum_{\vkp'} 
  \frac{e^{i \vkp' \cdot \vrp - i k_{\up z}(\varepsilon) z}}{\sqrt{v_{\uparrow z}(\varepsilon) W_x W_y}}
  \left( \begin{array}{c} 
  r_{\rm ee}(\varepsilon;\vkp',\vkp)  \\
  0 \\
  0 \\
  0 \end{array} \right)
  \\ && \nonumber \mbox{} +
  \sum_{\vkp'} 
  \frac{e^{i \vkp' \cdot \vrp - i k_{\up z}(-\varepsilon) z}}{\sqrt{v_{\uparrow z}(-\varepsilon) W_x W_y}}
  \left( \begin{array}{c} 
  0 \\
  0 \\
  r_{\rm he}(\varepsilon;\vkp',\vkp) \\
  0 \end{array} \right)
\end{eqnarray}
and
\begin{eqnarray}
  \Psi_{\varepsilon\vkp{\rm h}}(\vr) &=&
  \frac{e^{i \vkp \cdot \vrp - i k_{\up z}(\varepsilon) z}}{\sqrt{v_{\uparrow z}(-\varepsilon) W_x W_y}}
  \left(  \begin{array}{c} 0 \\ 0 \\ 1 \\ 0 \end{array} \right)
  \\ && \nonumber \mbox{}
  +
  \sum_{\vkp'} 
  \frac{e^{i \vkp' \cdot \vrp - i k_{\up z}(\varepsilon) z}}{\sqrt{v_{\uparrow z}(\varepsilon) W_x W_y}}
  \left( \begin{array}{c} 
  r_{\rm eh}(\varepsilon;\vkp',\vkp)  \\
  0 \\
  0 \\
  0 \end{array} \right)
  \\ && \nonumber \mbox{} +
  \sum_{\vkp'} 
  \frac{e^{i \vkp' \cdot \vrp + i k_{\up z}(-\varepsilon) z}}{\sqrt{v_{\uparrow z}(-\varepsilon) W_x W_y}}
  \left( \begin{array}{c} 
  0 \\
  0 \\
  r_{\rm hh}(\varepsilon;\vkp',\vkp) \\
  0 \end{array} \right),
\end{eqnarray}   
which represents a state with specified incoming electron-like or hole-like quasiparticle boundary conditions. Together, the amplitudes $r_{\rm ee}$, $r_{\rm eh}$, $r_{\rm he}$, and $r_{\rm hh}$ define the scattering matrix ${\cal S}$ of the HS interface,
\begin{equation}
  {\cal S}(\vkp',\vkp;\varepsilon) =
  \left( \begin{array}{cc} 
  r_{\rm ee}(\vkp',\vkp;\varepsilon)  &  r_{\rm eh}(\vkp',\vkp;\varepsilon)  \\
  r_{\rm he}(\vkp',\vkp;\varepsilon) &  r_{\rm hh}(\vkp',\vkp;\varepsilon)
  \end{array} \right).
\end{equation}
The scattering matrix is unitary. Particle-hole symmetry gives the further constraint\cite{kn:kupferschmidt2011}
\begin{eqnarray}
  r_{\rm ee}(\vkp',\vkp,\varepsilon) &=& r_{\rm hh}(-\vkp',-\vkp,-\varepsilon)^*, \nonumber \\
  r_{\rm eh}(\vkp',\vkp,\varepsilon) &=& r_{\rm he}(-\vkp',-\vkp,-\varepsilon)^*. \label{eq:ehsymmetry}
\end{eqnarray}

In the absence of the perturbation ${\cal V}$, ${\cal S}$ is found from a solution of the Bogoliubov-de Gennes equation (\ref{eq:BdG}) with ${\cal H} = {\cal H}_0$,
\begin{eqnarray}
  {\cal S}(\vkp',\vkp;\varepsilon) &=&
  \delta_{\vkp',\vkp}
  \left( \begin{array}{cc}
  \rho_{\up}(\vkp;\varepsilon) & 0 \\ 0 & \rho_{\up}(\vkp,-\varepsilon)^*
  \end{array} \right).
  \label{eq:S0}
\end{eqnarray}
Inclusion of the term ${\cal V}$ leads to a shift ${\cal S} \to {\cal S} + \delta {\cal S}$ of the scattering matrix, which can be calculated perturbatively by means of the Born series,
\begin{eqnarray}
  \label{eq:deltarhe}
  \delta r_{\rm he}(\vkp',\vkp;\varepsilon) &=&
  \frac{1}{i \hbar}
  \sum_{n=0}^{\infty}
  \,^{\rm A \!}\langle \varepsilon,\vkp',{\rm h}|
  \hat {\cal V} ({\cal G} {\cal V})^{n}
  | \varepsilon,\vkp,{\rm e} \rangle^{\rm R}, \nonumber \\
\end{eqnarray}
and similar expressions for $\delta r_{\rm ee}$, $\delta r_{\rm eh}$, and $\delta r_{\rm hh}$. Here the retarded scattering states $|\varepsilon,\vk_{\parallel},{\rm e}\rangle^{\rm R}$ and $|\varepsilon,\vk_{\parallel},{\rm h}\rangle^{\rm R}$ are solutions of the Bogoliubov-de Gennes equation (\ref{eq:BdG}) with ${\cal H} = {\cal H}_0$ and particle-like or hole-like incoming boundary conditions, respectively. The advanced scattering states $|\varepsilon,\vk_{\parallel},{\rm e}\rangle^{\rm A}$ and $|\varepsilon,\vk_{\parallel},{\rm h}\rangle^{\rm A}$ are solutions of the same equation, but with particle-like or hole-like outgoing boundary conditions. Explicitly, the wavefunction $\Psi^{\rm R}_{\varepsilon,\vk_{\parallel},{\rm e}}(\vr)$ of the electron-like scattering retarded state reads [compare Eqs.\ (\ref{eq:Psistandarde}) and (\ref{eq:S0})]
\begin{eqnarray}
  \label{eq:psi1}
  \Psi^{\rm R}_{\varepsilon \vkp {\rm e}}(\vr) &=&
  \frac{e^{i \vkp \cdot \vrp}}{\sqrt{v_{\uparrow z}(\varepsilon) W_x W_y}}
  \\ && \nonumber \mbox{} \times
  \left( \begin{array}{c} 
  e^{i k_{\uparrow z}(\varepsilon) z} +
  \rho_{\up}(\varepsilon) e^{-i k_{\uparrow z} (\varepsilon) z} \\
  0 \\
  0 \\
  \rho_{\dw}(-\varepsilon) e^{\kappa_{\downarrow z}(-\varepsilon) z}
  \end{array} \right), 
\end{eqnarray} 
whereas the wavefunction $\Psi_{\varepsilon,\vk_{\parallel},{\rm h}}(\vr)$ of the corresponding hole-like scattering state is obtained by particle-hole conjugation,
\begin{eqnarray}
  \label{eq:psi2}
  \Psi^{\rm R}_{\varepsilon\vkp{\rm h}}(\vr) &=&
  \frac{e^{i \vkp \cdot \vrp}}{\sqrt{v_{\uparrow z}(-\varepsilon) W_x W_y}}
  \\ && \nonumber \mbox{} \times
  \left( \begin{array}{c} 
  0 \\
  \rho_{\dw}(\varepsilon)^* e^{\kappa_{\downarrow z}(\varepsilon) z} \\ 
  e^{-i k_{\uparrow z}(-\varepsilon) z} + \rho_{\up}(-\varepsilon)^*
  e^{i k_{\uparrow z}(-\varepsilon) z} \\ 0
  \end{array} \right).
\end{eqnarray} 
The wavefunctions of the advanced scattering states are (recall $|\rho(\vkp,\varepsilon)| =1$)
\begin{eqnarray}
  \Psi^{\rm A}_{\varepsilon \vkp {\rm e}}(\vr) &=&
  \rho_{\up}(\varepsilon)^* \Psi^{\rm R}_{\varepsilon \vkp {\rm e}}(\vr),\nonumber \\
  \Psi^{\rm A}_{\varepsilon \vkp {\rm h}}(\vr) &=&
  \rho_{\up}(-\varepsilon) \Psi^{\rm R}_{\varepsilon \vkp {\rm h}}(\vr). 
  \label{eq:psiA}
\end{eqnarray}
Taken together, Eqs.\ (\ref{eq:deltarhe})--(\ref{eq:psiA}) and the expressions (\ref{eq:G01})--(\ref{eq:UU}) for the Green function ${\cal G}$ contain all information relevant for a calculation of the Andreev reflection amplitudes to arbitrary order in the perturbation ${\cal V}$.

\section{Andreev reflection amplitudes}\label{sec:perturbation}

We now describe the calculation of the Andreev reflection coefficients $r_{\rm he}(\varepsilon;\vk_{\parallel}',\vk_{\parallel}) $ and $r_{\rm eh}(\varepsilon;\vk_{\parallel}',\vk_{\parallel}) $ in the presence of a magnetization gradient and impurity scattering at the HS interface. All calculations are performed to first order in the magnetization gradient ({\em i.e.,} to first order in $l_{\rm d}^{-1}$). Without impurity scattering, one then finds Andreev reflection amplitudes that are diagonal in the wavevectors $\vkp$ and $\vkp'$ and linearly proportional to the excitation energy $\varepsilon$. This calculation was originally performed in Refs.\ \onlinecite{kn:beri2009,kn:kupferschmidt2011} and the result is briefly summarized in Sec.\ \ref{sec:3a} below. To first order in the impurity potential, we find Andreev reflections that are off-diagonal in $\vkp$ and $\vkp'$. These amplitudes remain finite at $\varepsilon=0$, but vanish for $\vkp = \vkp'$. In order to find the leading contribution to the diagonal amplitudes $r_{\rm he}(\varepsilon;\vk_{\parallel},\vk_{\parallel})$ and $r_{\rm eh}(\varepsilon;\vk_{\parallel},\vk_{\parallel})$ we thus need to go to second order in the impurity potential. The first-order and second-order calculations with respect to the impurity potential are given in Secs.\ \ref{sec:3b} and \ref{sec:3c}, where we restrict ourselves to the relevant cases $\varepsilon = 0$ and $\varepsilon=0$, $\vkp = \vkp'$, respectively.

\subsection{No impurity potential} \label{sec:3a}

To first order in the perturbation ${\cal V}$, only the magnetization gradient term $\hat V_{\rm m}$ contributes to the Andreev reflection amplitude. One finds \cite{kn:kupferschmidt2011}
\begin{eqnarray}
  r_{\rm he}(\varepsilon;\vk_{\parallel}',\vk_{\parallel}) &=&
  \delta_{\vk_{\parallel},\vk'_{\parallel}}
  \frac{i \varepsilon e^{-i \phi + i \varphi} \Delta_0 T_{\uparrow}(\vkp)}{4 k_{\up z} l_{\rm d}
  \sqrt{\Delta_0^2 - \varepsilon^2}}
  \nonumber \\ && \mbox{} \times
  \left[
  \frac{8 k_{\up z}}{
  \hbar v_{\dw z} (\kappa_{\dw z}^2 + k_{\up z}^2)}
  +
  \frac{T_{\uparrow}(\vkp)}{\sqrt{\Delta_0^2 - \varepsilon^2}}
  \right] \nonumber \\
  \label{eq:rhe10}
\end{eqnarray}
in the limit of a low transmission $T_{\up}(\vkp)$ of the half-metal--superconductor interface, where we kept the sub-leading term proportional to $T_{\up}(\vkp)^2$ because it appears with the (small) energy $\Delta_0$ in the denominator.\cite{footn3} (The divergence of Eq.\ (\ref{eq:rhe10}) for $|\varepsilon| \to \Delta_0$ is an artifact of the small-$T_{\up}$ expansion; Equation (\ref{eq:rhe10}) and Eq.\ (\ref{eq:Gld}) below are not valid for $|\varepsilon|$ in the immediate vicinity of $\Delta_0$.)

The essential feature of the Andreev reflection amplitude (\ref{eq:rhe10}) is that it vanishes at the Fermi level, $\varepsilon = 0$. As explained in the introduction, this is a consequence of the very special symmetries of the Hamiltonian (\ref{eq:BdG}) in the absence of impurity scattering. We now show by explicit calculation that impurity scattering, in combination with the magnetization gradient, gives rise to Andreev reflection into the half metal at zero energy. 

\subsection{First order in impurity potential}\label{sec:3b}

We now address the effect of impurities on the Andreev reflection amplitudes. We first consider a single impurity and postpone the discussion of the effect of a finite but low concentration of impurities Sec.\ \ref{sec:3d}. We choose our coordinates such, that the impurity is located at position $\vr_{\rm i} = (0,0,z_{\rm i})$ with $z_{\rm i} < 0$. The corresponding impurity potential then reads
\begin{equation}
  V_{\rm i}(\vr) = u_{\rm i} \delta(x) \delta(y) \delta(z - z_{\rm i}).
  \label{eq:Vimpone}
\end{equation}
With this choice of coordinates, the normal and Andreev reflection amplitudes are functions of the magnitudes $k_{\parallel}$ and $k_{\parallel}'$ of the initial-state and final-state wavevectors $\vkp$ and $\vkp'$ and the angle between these vectors only. In particular, with this impurity potential $\pi$-rotation symmetry around the $z$ axis is restored,
\begin{eqnarray}
  {\cal S}(\varepsilon;\vkp',\vkp) &=&
  {\cal S}(\varepsilon;-\vkp',-\vkp).
  \label{eq:inversion}
\end{eqnarray}

Evaluating the zero-energy Andreev reflection amplitude $r_{\rm he}$ to first order in $\hat V_{\rm m}$ and first order in the impurity potential $\hat V_{\rm i}$ of Eq.\ (\ref{eq:Vimpone}), we find a nonzero contribution to the Andreev reflection amplitude at zero energy,
\begin{widetext}
\begin{equation} \label{eq:reh11}
  r_{\rm he}(0;\vk_{\parallel}',\vk_{\parallel}) = 
  \frac{2 i u_{\rm i} e^{-i \phi + i \varphi}
  [ e^{\kappa_{\dw z}' z_{\rm i}}
  k_{\uparrow z} T_{\up}(\vkp') (\cos( k_{\uparrow z} z_{\rm i}) - e^{\kappa_{\dw z} z_{\rm i}})
  - \kappa_{\dw z} T_{\up}(\vkp) e^{\kappa_{\dw z} z_{\rm i}} \sin ( k_{\uparrow z}' z_{\rm i})
  ]}{\hbar l_\text{d} W_x W_y
  (\kappa_{\dw z}^2 + k_{\uparrow z}^2)
  \sqrt{v_{\up z} v_{\up z}'}}
  - \left( \vk_{\parallel} \leftrightarrow \vk_{\parallel}' \right),
\end{equation}
where, as before, we have given the result to leading order in the transparency $T_{\up}$ of the interface. [Unlike in the case of Eq.\ (\ref{eq:rhe10}) there are no terms of subleading order in $T_{\up}$ that come with small energy denominators.] The Andreev reflection amplitude $r_{\rm he}$ follows from the 
particle-hole symmetry relations (\ref{eq:ehsymmetry}). The above result vanishes if the impurity is located precisely at the superconductor interface. In that case, the leading contribution to the Andreev reflection amplitude is of higher order in the interface transparency. We have listed the corresponding expressions in App.\ \ref{sec:appendix0}.

The Andreev reflection amplitude (\ref{eq:reh11}) is odd under exchange of the initial and final momenta $\vkp$ and $\vkp'$. In particular, $r_{\rm he}(0;\vkp',\vkp)$ vanishes for $\vkp = \vkp'$. This can be seen from the following simple argument: For the calculation of the diagonal elements $r_{\rm he}(0;\vkp,\vkp)$ to first order in $V_{\rm m}$ and $V_{\rm i}$, there is no difference between an impurity potential of the form (\ref{eq:Vimpone}) and a potential that is invariant under translations parallel to the superconductor interface, $V_{\rm i}(\vr) = (W_x W_y)^{-1} u_{\rm i} \delta(z-z_{\rm i})$. For the latter potential, the general arguments of Ref.\ \onlinecite{kn:kupferschmidt2011} apply, from which it follows that $r_{\rm he}(0;\vkp,\vkp) = 0$.

Note that only impurities within a distance $\sim 1/\kappa_{\dw}$ of the minority electron wavefunction decay length from the interface contribute to the Andreev reflection amplitude $r_{\rm he}$. This is consistent with the picture that the Andreev reflection process involves an intermediate (evanescent) minority electron state, which is then converted into a majority state via a spin-flip process enabled by the combination of the impurity scattering and the magnetization gradient. Impurities at a larger distance from the superconductor interface do not contribute to the Andreev reflection amplitude. Their contribution to observable quantities, such as the conductance of an HS junction or the Josephson current in an SHS junction can be calculated using standard approaches, see, {\em e.g.}, Ref.\ \onlinecite{kn:beenakker1995}.

\subsection{Second order in impurity potential}\label{sec:3c}

A nonzero contribution to the diagonal amplitudes $r_{\rm eh}(0;\vkp,\vkp)$ can be expected in {\em second}-order perturbation theory in the impurity potential $V_{\rm i}$. Technically, it is most convenient to calculate the diagonal amplitudes $r_{\rm eh}(0;\vkp,\vkp)$ from the leading order ({\em first} order in $V_{\rm i}$) results for the off-diagonal amplitudes $r_{\rm eh}(0;\vkp',\vkp)$ of Eq.\ (\ref{eq:reh11}) and the first-order-in-$V_{\rm i}$ off-diagonal normal reflection amplitudes using the relation
\begin{equation}
  r_{\rm he}(0;\vkp,\vkp) = 
  - \sum_{\vkp' \neq \vkp} \frac{r_{\rm ee}(0;\vkp,\vkp')
  r_{\rm he}(0;\vkp,\vkp')}{r_{\rm ee}(0;\vkp,\vkp)},
  \label{eq:opticaltheorem}
\end{equation}
which is obtained upon combining unitarity of the scattering matrix, the particle-hole symmetry relations (\ref{eq:ehsymmetry}), and the symmetry relations (\ref{eq:inversion}). For the off-diagonal normal reflection amplitude $r_{\rm ee}(0;\vkp',\vkp)$ we find
\begin{eqnarray}
  r_{\rm ee}(0;\vkp',\vkp) &=& \frac{4 i u_{\rm i} \sin(k_{\up z} z_{\rm i})
  \sin(k_{\up z}' z_{\rm i})}{\hbar W_x W_y \sqrt{v_{\up z} v_{\up z}'}},
  \label{eq:ree10}
\end{eqnarray}
to leading (zeroth) order in the interface transparency. 
To leading order in the interface transparency, the diagonal normal reflection coefficient $r_{\rm ee}(0;\vkp,\vkp) = -1$. Because $r_{\rm he}(0;\vkp,\vkp')$ vanishes for $|\vkp| = |\vkp'|$, the summation in Eq.\ (\ref{eq:opticaltheorem}) is dominated by terms in which $|\vkp| - |\vkp'| \sim k_{\up}$.
The expression for the diagonal second-order-in-$V_{\rm i}$ Andreev reflection amplitude that we find upon substituting Eqs.\ (\ref{eq:reh11}) and (\ref{eq:ree10}) into Eq.\ (\ref{eq:opticaltheorem}) is too lengthy to be reported in full, but the result takes a simple form in the limit $v_{\up} \ll \min(v_{\downarrow},v_{\rm S}) \ll \max(|w|,v_{\rm S})$, corresponding to a large mismatch of Fermi velocities in H and S,
\begin{equation}
  r_{\rm he}(0;\vkp,\vkp) =
  \frac{2 u_{\rm i}^2 e^{-i (\phi - \varphi) + \kappa_{\dw} z_{\rm i}} z_{\rm i}^2 k_{\up}^3 ( 5 k_{\up z}^2 - 3 k_{\up}^2 )  T_{\up}(\vkp)}{45 \hbar^2 W_x W_y \kappa_{\dw}^2 v_{\dw}^2 l_d \pi}
  \left[ 6(1 - e^{\kappa_{\dw} z_{\rm i}})
  - 6 \kappa_{\dw} z_{\rm i}
  + 3 (\kappa_{\dw} z_{\rm i})^2
  + (\kappa_{\dw} z_{\rm i})^3 \right].
  \label{eq:rhe12}
\end{equation}
\end{widetext}
Equation (\ref{eq:rhe12}) is valid for arbitrary impurity locations $\vr_{\rm i}$. Only impurities within a distance $\sim 1/\kappa_{\dw}$ contribute to the Andreev reflection amplitude of Eq.\ (\ref{eq:rhe12}). Note that the first order amplitude (\ref{eq:reh11}) and the second-order amplitude (\ref{eq:rhe12}) are both proportional to the same power of the interface transparency $T_{\up}(\vkp)$. 

\subsection{Finite density of impurities}\label{sec:3d}

We now consider a finite but low density $n_{\rm imp}$ of impurities in the immediate vicinity of the surface, for which the potential strength $u_{j}$ is itself a random variable with zero mean and with variance $\langle u_{j}^2 \rangle$. In view of the applications of the next section, we are interested in the ensemble averages $\langle r_{\rm he}(0;\vkp',\vkp) \rangle$ and $\langle |r_{\rm he}(0;\vkp',\vkp)|^2 \rangle$ with respect to the disorder at the interface, to lowest (first) order in $n_{\rm imp}$. The single-impurity results derived above are sufficient for this calculation, since interference effects between Andreev reflection processes that are enabled by scattering off different impurities can be neglected to this order in the impurity density $n_{\rm imp}$. (They give a contribution proportional to $n_{\rm imp}^2$.)

Thus proceeding, we find that the average reflection amplitude $\langle r_{\rm he}(0;\vkp',\vkp) \rangle$ is nonzero only if $\vkp' = \vkp$ (because translation invariance along the interface is restored upon taking the ensemble average), 
\begin{eqnarray}
  \langle r_{\rm he}(0;\vkp,\vkp) \rangle &=&
  -\delta_{\vkp',\vkp} e^{-i (\phi - \varphi)} n_{\rm imp} \langle \sigma \rangle   T_{\up}(\vkp)
  \nonumber \\ && \mbox{} \times
  \frac{
  k_{\up}^3 (5 k_{\up z}^2 - 3 k_{\up}^2)}{15 \kappa_{\dw}^7 l_{\rm d}}. \label{eq:ravg}
\end{eqnarray}
where $\langle \sigma \rangle = m^2 \langle u_j^2 \rangle/\pi \hbar^4$ is the mean scattering cross section of the impurities, see Eq.\ (\ref{eq:sigma}). The mean square Andreev reflection amplitude is
\begin{eqnarray}
  \langle \langle |r_{\rm he}(0;\vkp',\vkp)|^2 \rangle &=& n_{\rm imp} \langle \sigma \rangle  T_{\up}(\vkp') T_{\up}(\vkp) \nonumber \\ && \mbox{} \times
  \frac{2185 \pi (k_{\up z}^2 - k_{\up z}'^2)^2}{648 \kappa_{\dw}^9 l_{\rm d}^2 W_x W_y}.
\end{eqnarray}
Both results are for the limit $v_{\up} \ll \min(v_{\downarrow},v_{\rm S}) \ll \max(|w|,v_{\rm S})$.

It is an interesting question, what the effect of impurities is on the symmetry of the superconducting correlations in the half metal. In the quasiclassical Green function approach, the superconducting correlations in the half metal are described with the help of the anomalous Green function $f(i \Omega;\vk,\vr)$, $\Omega$ being the Matsubara frequency, which is related to the Andreev reflection amplitudes as\cite{kn:kupferschmidt2011}
\begin{eqnarray}
  f(i \Omega;\vk,\vr) &=&
  2 e^{-2 |\Omega| (\vr \cdot \ve_z)/\hbar v_{\up z}} \nonumber \\ && \mbox{}
  \times
  \left\{ \begin{array}{ll}
  r_{\rm eh}(i \Omega;\vkp,\vkp) & \mbox{$k_z < 0$, $\Omega > 0$,} \\
  -r_{\rm he}(-i \Omega;\vkp,\vkp)^* & \mbox{$k_z > 0$, $\Omega < 0$},\\
  0 & \mbox{otherwise}, \end{array} \right.\nonumber \\
  \label{eq:f}
\end{eqnarray}
where $\vk$ is a wavevector with $k = k_{\up}$. Without impurities at the interface, one thus finds that the spin-active interface gives rise to correlations of predominantly odd-frequency $s$-wave type, with an anomalous Green function $f \propto \Omega$ for small frequencies.\cite{kn:eschrig2009,kn:kupferschmidt2011} We find from Eqs.\ (\ref{eq:ravg}) and (\ref{eq:f}) that the presence of impurities at the interface does not change this fundamental symmetry of the induced superconducting order, but it does change the asymptotic low-frequency dependence of $f$, $f$ being proportional to $\mbox{sign}\,(\Omega)$ for small $\Omega$. This enhancement of odd frequency $s$-wave superconducting correlations at small frequencies is a unique signature of the impurity-assisted spin-triplet proximity effect in half metals.

\section{Applications}\label{sec:applications}

\subsection{Conductance}\label{sec:conductance}

The conductance of a half-metal--superconductor interface is given by the expression \cite{kn:blonder1982,kn:beenakker1995}
\begin{equation}
  G(V) = \frac{2 e^2}{h} \sum_{\vkp,\vkp'}
  |r_\text{he}(\varepsilon=eV;\vkp',\vkp)|^2,
  \label{eq:GNS}
\end{equation}
Without impurities, $r_{\rm he}$ is given by Eq.\ (\ref{eq:rhe10}) if there is a magnetization gradient perpendicular to the interface, and $G$ is proportional to $V^2$ at low voltage,\cite{kn:beri2009,kn:kupferschmidt2011}
\begin{eqnarray}
  G(V) &=& \frac{e^2 (eV)^2 W_x W_y \Delta_0^2 T_{\up}(0)^2}{2 \pi^2 \hbar l_{\rm d}^2 (\Delta_0^2 - (eV)^2)}
  \left[ \frac{k_{\up}^2}{\hbar^2 v_{\dw}^2 \kappa_{\dw}^4}
  \right. \label{eq:Gld} \nonumber \\ && \left. \mbox{}
  + \frac{k_{\up} T_{\up}(0)}{4 \hbar v_{\dw}\kappa_{\dw}^2 \sqrt{\Delta_0^2 - (eV)^2}}
  + \frac{T_{\up}(0)^2}{64(\Delta_0^2 - (eV)^2)} \right], \nonumber \\
\end{eqnarray}
where $T_{\up}(0)$ is the normal-state transmission coefficient of the clean HS interface for perpendicular incidence ($\vkp = 0$), and we took the limit $v_{\up} \ll \min(v_{\downarrow},v_{\rm S}) \ll \max(|w|,v_{\rm S})$.

For low bias, an impurity at position $\vr_{\rm i} = z_{\rm i} \ve_z$ {\em increases} the conductance by a finite amount, which is found by substituting Eq.\ (\ref{eq:reh11}) into Eq.\ (\ref{eq:GNS}). Because of space limitations, we here report the resulting expression for the limit $v_{\up} \ll \min(v_{\downarrow},v_{\rm S}) \ll \max(|w|,v_{\rm S})$ only,
\begin{eqnarray}
  \delta G &=&
  \frac{2 e^2 \sigma_{\rm i} T_{\up}^2(0) k_{\up}^8}{4725 \pi^2 \hbar l_{\rm d}^2 \kappa_{\dw}^8} e^{2 \kappa_{\dw} z_{\rm i}}  \label{eq:deltaG}
  \\ && \mbox{} \times
  \left[6 - 6 e^{\kappa_{\dw} z_{\rm i}} 
  - 6 \kappa_{\dw} z_{\rm i}
  + 3 (\kappa_{\dw} z_{\rm i})^2
  + (\kappa_{\dw} z_{\rm i})^3 \right]^2.   \nonumber 
\end{eqnarray}
where $\sigma_{\rm i}$ is the scattering cross section of the impurity, see Eq.\ (\ref{eq:sigma}). For a finite but small concentration $n_{\rm imp}$ of impurities near the superconductor interface one may add these contributions to the conductance, giving a finite conductance at zero bias,
\begin{equation}
  G(0) = \frac{437 e^2 n_{\rm imp} \langle \sigma \rangle W_x W_y T_{\up}(0)^2 k_{\up}^8}{34020 \pi^2 \hbar \kappa_{\dw}^9 l_{\rm d}^2},
  \label{eq:Gimp}
\end{equation}
where $\langle \sigma \rangle$ is the mean scattering cross section of the impurities. 
At finite bias, the conductance is the sum of Eqs.\ (\ref{eq:Gld}) and (\ref{eq:Gimp}).

The impurity-assisted contribution to the conductance (\ref{eq:Gimp}) dominates over the conductance (\ref{eq:Gld}) of a clean HS interface in the limit of low bias voltages,
\begin{equation}
  \left(\frac{eV}{|\mu_{\text{H}\dw}|} \right)^2 \lesssim  
  \frac{n_{\rm imp} \langle \sigma \rangle}{\kappa_{\dw}} 
  \left( \frac{k_{\up}}{\kappa_{\dw}} \right)^6,
  \label{eq:Gineq}
\end{equation}
where $\mu_{\text{H}\dw} = -\hbar \kappa_{\dw} v_{\dw}/2$ is the excitation gap for minority quasiparticles in the half metal, see Eq.\ (\ref{eq:kappadown}). This condition is independent of the normal-state transmission coefficient $T_{\up}(0)$ and the magnetization gradient $l_{\rm d}^{-1}$, since the interface conductance $G$ is proportional to $T_{\up}(0)^2 l_{\rm d}^{-2}$ for a clean interface as well as for an interface with impurity scattering. (Here we took the first term in Eq.\ (\ref{eq:Gld}) as the basis for our comparison, which is the leading contribution to the conductance of a clean interface in the limit of a small interface transparency $T_{\up}(0)$.) The fraction on the left hand side of Eq.\ (\ref{eq:Gineq}) is the total cross section of all impurities within a layer of width $1/\kappa_{\dw}$ adjacent to the HS interface, per unit interface area.

\subsection{Josephson current}\label{sec:josephsoncurrent}

\begin{figure}
  \centering
  \includegraphics[width = 0.95 \linewidth]{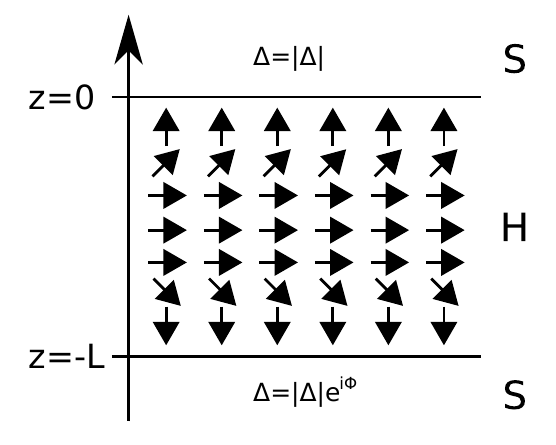}
  \caption{SHS junction of length $L$ with two spin-active interfaces.}
  \label{fig:SHS-matrix2}
\end{figure}

Next, we consider a superconductor--half-metal--superconductor junction, in which the central half-metallic segment is clean, with the possible exception of the presence of impurities with density $n_{\rm imp}$ near the two interfaces. We choose coordinates such that the two interfaces are at $z=-L$ and $z=0$, see Fig.\ \ref{fig:SHS-matrix2}.
We consider the case that the magnetization gradient is perpendicular to the superconductor interface at both interfaces, with equal azimuthal angles $\varphi$, and that the magnetization gradients are equal in magnitude and opposite in direction. 
We restrict our analysis to the so-called ``long junction'' limit $E_{L} \ll \Delta_0$, where
\begin{equation}
  E_L = \frac{\hbar v_{\up}}{2 \pi L}
\end{equation}
is the Thouless energy of the junction, and we consider temperatures in the range $E_L \ll k_{\rm B}T \ll \Delta_0$.
In this temperature regime, the Josephson current is quadratic in the Andreev reflection amplitudes,\cite{kn:brouwer1997e}
\begin{eqnarray}
  I &=& - \frac{4 e k_{\rm B} T}{\hbar}
  \frac{d}{d \phi} \sum_{\vkp, \vkp'}
  e^{-\pi k_{\rm B} T L/\hbar v_{\up z} - \pi k_{\rm B} T L/\hbar v_{\up z}'}
  \nonumber \\ && \mbox{} \times
  \mbox{Re}\,
  r_{\rm eh}(i \pi k_{\rm B} T;\vkp,\vkp')
  \tilde r_{\rm he}(i \pi k_{\rm B} T;\vkp',\vkp),
  \label{eq:I}
\end{eqnarray}
where $k_{\rm B}$ is the Boltzmann constant, $T$ temperature, $\phi$ the phase difference between the two superconductors.
The amplitudes $r_{\rm eh}$ and $\tilde r_{\rm he}$ describe Andreev reflection at the interfaces at $z=0$ and $z=-L$, respectively.

Each Andreev reflection amplitude that appears in Eq.\ (\ref{eq:I}) is the sum of two contributions: A contribution for the clean half-metal--superconductor interface, given in Eq.\ (\ref{eq:rhe10}), and a contribution from impurity-mediated Andreev reflection. Upon taking the average over the ensemble of impurities, the impurity-mediated contributions two factors $r_{\rm eh}$ and $\tilde r_{\rm he}$ can be averaged separately, as they refer to two different interfaces. After taking the ensemble average, translation symmetry along the interface is restored, and the ensemble-averaged amplitudes $\langle r_{\rm eh}(i \pi k_{\rm B} T;\vkp',\vkp) \rangle$ and $\langle \tilde r_{\rm he}(i \pi k_{\rm B} T;\vkp',\vkp) \rangle$ are zero except if $\vkp = \vkp'$, see Eq.\ (\ref{eq:ravg}).
We then find, again in the limit $v_{\up} \ll \min(v_{\downarrow},v_{\rm S}) \ll \max(|w|,v_{\rm S})$, that
\begin{eqnarray} \label{eq:Iresult}
  I &=& - \frac{e \pi W_x W_y T_{\up}(0)^2 E_L}{1800 \hbar l_{\rm d}^2}
  e^{-k_{\rm B} T/E_L}
  \sin \phi  \\ && \mbox{} \times \nonumber
  \left( \frac{15 k_{\rm B} T T_{\up}(0)}{\Delta_0}
  + \frac{120 k_{\rm B} T k_{\up}}{\hbar v_{\dw} \kappa_{\dw}^2}
  - \frac{8 k_{\up}^6 \langle \sigma \rangle n_{\rm imp}}{\pi \kappa_{\dw}^7}
  \right)^2.
\end{eqnarray}
Upon lowering the temperature, Eq.\ (\ref{eq:Iresult}) remains valid as long as $k_{\rm B} T \gtrsim E_L$, and the Josephson current saturates for temperatures $k_{\rm B} T \sim E_L$.

Comparing the Josephson current for a clean interface and the impurity-assisted contribution, we find that the impurity-assisted contribution dominates if
\begin{equation}
  \frac{n_{\rm imp} \langle \sigma \rangle}{\kappa_{\dw}} 
  \left( \frac{k_{\up}}{\kappa_{\dw}} \right)^5
  \gtrsim \frac{\max(k_{\rm B} T,E_L)}{|\mu_{\text{H}\dw}|},
\end{equation}  
where we again took the current in the clean case in the limit of a low transparency of the HS interface. Again, this condition is independent of $T_{\up}$ and $l_{\rm d}$, because the dependence of the Josephson current on these parameters is the same (proportional to $T_{\up}^2 l_{\rm d}^{-2}$) for clean and dirty interfaces.

\section{Conclusion}\label{sec:summary}

We have verified through explicit calculation that impurities in the vicinity of a half-metal--superconductor (HS) junction with a magnetization gradient perpendicular to the interface (spin-active interface) qualitatively change the dependence of the Andreev reflection amplitude $r_{\rm he}$ on the quasiparticle energy $\varepsilon$ and, hence, the bias dependence of the interface conductance $G$ and the temperature dependence of the Josephson current $I$ of a superconductor--half-metal--superconductor (SHS) junction. Without impurities, one has $r_{\rm he} \propto \varepsilon$ at low energy.\cite{kn:beri2009} In contrast, impurities give rise to a nonzero value of $r_{\rm he}$ for $\varepsilon=0$. Interestingly, the zero-bias conductance of a HS junction and the low-temperature Josephson current of an SHS junction increase upon increasing the impurity concentration at the interface.

Although it was to be expected, based on general symmetry considerations, that impurity scattering leads to a nonzero Andreev reflection {\em probability} $|r_{\rm he}|^2$, we found the remarkable result that a nonzero amplitude $\langle r_{\rm he} \rangle$ remains after taking an average over impurity locations and potentials. It is this coherent impurity-assisted Andreev reflection that is responsible for the increase in the low-temperature Josephson current and the superconducting correlations at low frequency that we predict. The nonzero average persists if an average over the Fermi surface is performed.

There is an important difference between the impurities in the immediate vicinity of the interface we considered here, and impurities at a larger distance into the half metal. Only impurities in the half metal that are within a minority-electron wavefunction decay length from the interface may enhance Andreev reflection. Within a quasiclassical approach, such impurities are considered as part of the interface, and they are an integral part of the boundary conditions that have to be applied at the half-metal--superconductor interface. Impurities that are located farther into the half metal lead to scattering of quasiparticles before and after Andreev reflection, but such impurities are not involved in the Andreev reflection process itself. Their effect can be treated with standard methods from quasiclassics or the scattering matrix approach, which combine normal-state propagation inside the half metal with the Andreev-reflection boundary conditions at the half-metal--superconductor interface.\cite{kn:beenakker1995,kn:eschrig2007} In any case, since the induced superconducting order is of $s$-wave type for both clean and disordered spin-active interfaces, impurities away from the interface will only have a minor effect on the superconducting order induced in the half metal. 

\acknowledgments

We gratefully acknowledge discussions with Joern Kupferschmidt and Mathias Duckheim. This work is supported by the Alexander von Humboldt Foundation in the framework of the Alexander von Humboldt Professorship program, endowed by the Federal Ministry of Education and Research, and by SPP 1538 of the DFG.

\appendix

\section{Scattering states}
\label{sec:appa}

The full expression for the column vectors $V_{\uparrow}$ and $V_{\downarrow}$ appearing in the Green function ${\cal G}_z$ is
\begin{widetext}
\begin{eqnarray}
  V_{\uparrow} &=& \frac{1}{\hbar v_{\uparrow z}}
  \left( \begin{array}{c} 0 \\
  \rho_{\dw}(\varepsilon)^* e^{i k_{\uparrow z}(-\varepsilon) z' + \kappa_z(\varepsilon) z} \\ 
  e^{-i k_{\uparrow z}(-\varepsilon)|z-z'|} + \rho_{\up}(-\varepsilon)^* e^{i k_{\uparrow z}(-\varepsilon)(z+z')} \\
0 \end{array} \right), \nonumber \\
  V_{\downarrow} &=& \frac{1}{(-i) v_{\downarrow z}}
  \left( \begin{array}{c}
  \tau_{\up}(\varepsilon)^* e^{-i k_{\uparrow z}(\varepsilon)  z + \kappa_z(-\varepsilon) z'} \\ 0 \\ 0 \\
  e^{-\kappa_z(-\varepsilon)|z-z'|} + \tau_{\dw}(-\varepsilon)^* e^{\kappa_z(-\varepsilon)(z+z')} \end{array} \right). \\
\end{eqnarray} 
The detailed expressions for the coefficients $\rho_{\sigma}$ and $\tau_{\sigma}$ of Eq.\ (\ref{eq:UU}) are
\begin{eqnarray}
  \rho_{\up}(\varepsilon) &=&
  \frac{-v_{{\rm S}z} (v_{\downarrow z} - i v_{\uparrow z}) \varepsilon
   + (v_{{\rm S}z}^2 + (v_{\downarrow z} + 2 w) (i v_{\uparrow z} + 2 w)) \sqrt{\Delta_0^2 - \varepsilon^2}}
  {v_{{\rm S}z} (v_{\downarrow z} + i v_{\uparrow z})  \varepsilon
   - (v_{{\rm S}z}^2 + (v_{\downarrow z} + 2 w) (-i v_{\uparrow z} + 2 w)) \sqrt{\Delta_0^2 - \varepsilon^2}}, \\
  \rho_{\dw}(\varepsilon) &=& - \frac{2 i e^{-i \phi} v_{{\rm S}z} v_{\uparrow z} \Delta_0}
  {v_{{\rm S}z} (v_{\downarrow z} + i v_{\uparrow z})  \varepsilon
      + (v_{{\rm S}z}^2 + (v_{\downarrow z} + 2 w) (-i v_{\uparrow z} + 2 w)) \sqrt{\Delta_0^2 - \varepsilon^2}}, \\
  \tau_{\dw}(\varepsilon) &=&
  \frac{v_{{\rm S}z} (v_{\downarrow z} + i v_{\uparrow z}) \varepsilon
     - (v_{{\rm S}z}^2 - (v_{\downarrow z} - 2 w) (i v_{\uparrow z} + 2 w)) 
  \sqrt{\Delta_0^2 - \varepsilon^2}}
  {v_{{\rm S}z} (v_{\downarrow z} - i v_{\uparrow z}) \varepsilon
     + (v_{{\rm S}z}^2 + (v_{\downarrow z} + 2 w) (i v_{\uparrow z} + 2 w)) \sqrt{\Delta_0^2-\varepsilon^2}}, \\
  \tau_{\up}(\varepsilon) &=&
  \frac{2 e^{-i \phi}
    v_{\downarrow z} v_{{\rm S}z} \Delta_0}{v_{{\rm S}z} (v_{\downarrow z} - i v_{\uparrow z}) \varepsilon - (v_{{\rm S}z}^2 + (v_{\downarrow z} + 2 w) (i v_{\uparrow z} + 2 w)) \sqrt{\Delta_0^2-\varepsilon^2}}.
\end{eqnarray} 
\end{widetext}

\section{Impurities precisely at the HS interface}
  \label{sec:appendix0}

If all impurities are located precisely at the HS interface ({\em i.e.}, $z_{\rm i} = 0$ for all impurities), the impurity-assisted Andreev reflection amplitude vanishes to the order in the interface transparency that was required to derive Eqs.\ (\ref{eq:reh11}) and (\ref{eq:rhe12}) of the main text. In this appendix we collect the main results of this article for the case $z_{\rm i} = 0$.

For impurities located at the HS interface it is not possible to eliminate the parameters $w$ and $v_{{\rm S}z}$ in favor of the normal-state transmission $T_{\up}(\vkp)$ of the HS interface only. In addition, we need the imaginary part $\mbox{Im}\, r_{\up}(\vkp)$ of the reflection amplitude for majority electrons incident on the HS interface from the half metal,
\begin{equation}
  r_{\up}(\vkp) = \frac{-2 i w - v_{{\rm S}z} + v_{\up z}}{2 i w + v_{{\rm S}z} + v_{\up z}}.
\end{equation}
The transmission probability $T(\vkp)$ for majority electrons incident at the half-metal--superconductor interface with the superconductor in the normal state is $T_{\uparrow}(\vk) = 1 - |r_{\up}(\vkp)|^2$.

For the first-order-in-$\hat V_{\rm i}$ contribution to the Andreev reflection amplitude one then finds
\begin{widetext}
\begin{eqnarray}
  r_{\rm he}(0;\vk_{\parallel}',\vk_{\parallel}) &=& 
  \frac{i u_{\rm i} e^{-i \phi + i \varphi}
  \kappa_{\dw z}' 
  [T_{\up}(\vkp') \mbox{Im}\, r_{\up}(\vkp) +
   T_{\up}(\vkp) \mbox{Im}\, r_{\up}(\vkp')]}
  {\hbar l_{\rm d} W_x W_y (\kappa_{\dw z}'^2 + k_{\up z}'^2) 
  \sqrt{v_{\up z} v_{\up z}'}}
  - \left( \vk_{\parallel} \leftrightarrow \vk_{\parallel}' \right),
  \label{eq:reh11b}
\end{eqnarray}
to leading order in the interface transparency.
The normal reflection amplitude is, again to first order in the impurity potential and to leading order in the interface transparency,
\begin{eqnarray}
  r_{\rm ee}(0;\vkp',\vkp) &=& 
  - \frac{i u_{\rm i} [T_{\up}(\vkp) T_{\up}(\vkp') - 4 \mbox{Im}\, r_{\up}(\vkp) \mbox{Im}\, r_{\up}(\vkp')]}{4 \hbar W_x W_y \sqrt{v_{\up z} v_{\up z}'}}.
  \label{eq:ree10b}
\end{eqnarray}
For the second-order-in-$V_{\rm i}$ diagonal Andreev reflection amplitude one finds
\begin{eqnarray}
  r_{\rm he}(0;\vkp,\vkp) &=&
  \frac{u_{\rm i}^2 e^{-i (\phi - \varphi)}
  k_{\up}^3 (\mbox{Im}\, r_{\up}(\vkp))
  T_{\up}(\vkp)
  [T_{\up}(\vkp)^2 - 4 (\mbox{Im}\, r_{\up}(\vkp))^2] (5 k_{\up z}^2 - 3 k_{\up}^2)}{60 W_x W_y \hbar^2 \kappa_{\dw} v_{\dw}^2 l_{\rm d} \pi k_{\up z}^3},
  \label{eq:rhe12b}
\end{eqnarray}
where we took the limit $v_{\up} \ll \min(v_{\downarrow},v_{\rm S}) \ll \max(|w|,v_{\rm S})$. In the same limit one finds that a single impurity at the HS interface increases the zero-bias conductance of a HS junction by the amount
\begin{eqnarray}
  \delta G &=&
  \frac{8 e^2 \langle \sigma \rangle T_{\up}(0)^2 [\mbox{Im}\, r_{\up}(0)]^2 k_{\up}^6}{525 \pi^2 \hbar \kappa_{\dw}^6 l_{\rm d}^2}.
\end{eqnarray}
Finally, for impurities located precisely at the superconductor interface with surface density $n_{\rm imp, 2d}$, we find the Josephson current
\begin{eqnarray}
  I &=& - \frac{e \pi W_x W_y T_{\up}(0)^2 E_L}{1800 \hbar l_{\rm d}^2} \nonumber
  e^{-k_{\rm B} T/E_L}
  \sin \phi  \\ && \mbox{} \times 
  \left( \frac{15 k_{\rm B} T T_{\up}(0)}{\Delta_0}
  + \frac{120 k_{\rm B} T k_{\up}}{\hbar v_{\dw} \kappa_{\dw}^2}
  - \frac{2 k_{\up}^3 \langle \sigma \rangle n_{\rm imp, 2d}
  (\mbox{Im}\, r_{\up}(0))
  [T_{\up}(0)^2 - 4 (\mbox{Im}\, r_{\up}(0))^2]}{\pi \kappa_{\dw}^3}
  \right)^2.
\end{eqnarray}

\section{Ideal interface} \label{sec:appendixideal}

Here, we analyze the case of an ideal half-metal--superconductor interface with perfect transmission, keeping the (non-self-consistent) step-function model for the order parameter $\Delta(\vr)$. Perfect transparency $T_{\uparrow}(\vk_{\parallel}) = 1$ at the NS interface is achieved by setting $w = 0$ and $v_{\rm S} = v_{\up}$. In analogy to Eq.\ (\ref{eq:reh11}), the leading contribution to the impurity-assisted Andreev reflection amplitude,
\begin{eqnarray}
  r_{\rm he}(0;\vkp',\vkp) &=&
  \frac{- 4 i u_{\rm i} e^{-i \phi + i \varphi}}{l_{\rm d} \hbar W_x W_y
  (k_{\up z}^2 + \kappa_{\dw z}^2)(k_{\up z} + i \kappa_{\dw z})
  (k_{\up z}' - i \kappa_{\dw z}')
  \sqrt{v_{\up z}' v_{\up z}}}
  \left[\left\{- 2 e^{\kappa_{\dw z}' z_{\rm i}} k_{\up z}' k_{\up z}
  [k_{\up z} \cos(k_{\up z} z_{\rm i}) 
  + \kappa_{\dw z} \sin(k_{\up z} z_{\rm i})]
  \right. \right. \nonumber \\ && \left. \left. \mbox{} 
  + 2 e^{\kappa_{\dw z} z_{\rm i}} k_{\up z} \kappa_{\dw z}
  [k_{\up z}' \sin(k_{\up z}' z_{\rm i}) -
  \kappa_{\dw z}' \cos(k_{\up z}' z_{\rm i})]
  - e^{\kappa_{\dw z} z_{\rm i} + \kappa_{\dw z}' z_{\rm i}}
  k_{\up z}' (\kappa_{\dw z}^2 - 3 k_{\up z}^2)
 \right\} - \left( \vkp \leftrightarrow \vkp' \right) \right],
\end{eqnarray}
vanishes for $\vkp = \vkp'$. For the first-order-in-$V_{\rm i}$ normal reflection amplitude, we find in analogy to Eq.\ (\ref{eq:ree10}), that
\begin{equation}
r_{\rm ee}(0;\vkp',\vkp) = \frac{-4 i u_{\rm i} e^{i \phi - i \varphi} \left[ k_{\uparrow z} k_{\uparrow z}' e^{x (\kappa_{\downarrow z}+\kappa_{\downarrow z}')}-(\kappa_{\downarrow z} \cos (x k_{\uparrow z})-k_{\uparrow z} \sin (x k_{\uparrow z})) (\kappa_{\downarrow z}' \cos (x k_{\uparrow z}')-k_{\uparrow z}' \sin (x k_{\uparrow z}'))\right]}{\hbar W_x W_y \sqrt{v_{\uparrow z} v_{\uparrow z}'} (k_{\uparrow z}-i \kappa_{\downarrow z}) (k_{\uparrow z}'-i \kappa_{\downarrow z}')}.
\end{equation}
The remaining results in this appendix are for the limit $v_{\rm S} = v_{\up} \ll v_{\downarrow}$, corresponding to a large mismatch of the (minority) Fermi velocities in the half metal and the superconductor.
For the diagonal contribution to $r_{\rm he}$, we then find
\begin{eqnarray} 
r_{\rm he}(0;\vkp,\vkp) &=& \frac{4 u_{\rm i}^2 e^{-i \phi+i \varphi} }{\hbar^2 W_x W_y l_\text{d} k_{\uparrow z}} \frac{k_\uparrow (k_\uparrow - 2 k_{\uparrow z})}{\pi v_{\downarrow}^2}  e^{z_i \kappa_{\downarrow z}} (2 - e^{z_i \kappa_{\downarrow z}}) \, .
\end{eqnarray} 
Taking the average in impurity position in the half-metal (\ref{eq:ravg}) leads us to
\begin{eqnarray}
\langle r_{\rm he}(0;\vkp,\vkp) \rangle &=&  \frac{6 \langle \sigma \rangle n_\text{imp} e^{-i \phi+i \varphi} k_\uparrow (k_\uparrow - 2 k_{\uparrow z})}{ l_\text{d} \kappa_\downarrow^3 k_{\uparrow z}} \, .
\end{eqnarray}
As described in subsection \ref{sec:conductance}, we calculate the leading correction to the subgap conductance caused by a single impurity in the half metal,
\begin{eqnarray}
\delta G &=& \frac{2 e^2}{h} \frac{2 \sigma_i}{\pi l_\text{d}^2} e^{2 \kappa_{\dw z} z_{\rm i}} \frac{(e^{\kappa_{\dw z} z_{\rm i}} - 2)^2}{3} \frac{k_{\uparrow}^4}{\kappa_{\uparrow z}^4} \, ,
\end{eqnarray}
and find, after taking the average in impurity position,
\begin{eqnarray}
G(0) &=& \frac{2 e^2}{h} \frac{11 \langle \sigma \rangle n_\text{imp}  W_x W_y}{18 \pi l_\text{d}^2} \frac{k_{\uparrow}^4}{\kappa_{\uparrow z}^5} \, .
\end{eqnarray}
Finally, for the impurity-assisted Josephson current in the long junction limit for high temperatures $E_L \ll k_{\rm B}T \ll \Delta_0$ we find
\begin{align}
I =& \frac{16 e \, v_\uparrow}{L} \; \frac{W_x W_y k_\uparrow^2}{4 \pi^2} \left(\frac{1}{k_\uparrow l_\text{d}} \right)^2  \frac{k_\uparrow^4}{\kappa_\downarrow^4} \cdot \sin(\phi) \cdot e^{-k_\text{B} T / E_\text{L}} \left(\frac{2 \pi k_\text{B} T}{\Delta_0} - \frac{3 n_\text{imp} \langle \sigma \rangle}{\kappa_\downarrow}  \right)^2 \, .\label{JCur}
\end{align}
\end{widetext}


\begin{thebibliography}{37}
\bibitem{footn2}The Green function ${\cal G}_z(\varepsilon;\vk_{\parallel};z,z')$ is a solution of the equation
$$
  (\varepsilon - {\cal H}_z){\cal G}_z(\varepsilon;\vk_{\parallel};z,z') =
  \delta(z-z'),
$$
where ${\cal H}_z$ is obtained from ${\cal H}$ by the replacements ${\vnabla^2 \to \partial^2/\partial z^2}$ and ${\mu_{\sigma} \to \mu_{\sigma} - \hbar^2 k_{\parallel}^2/2m}$.
\end{thebibliography}
\end{document}